\documentclass[9pt, twocolumn]{article}

\usepackage{geometry}
\usepackage{graphicx}
\usepackage{dcolumn}
\usepackage{bm}

\usepackage{amsmath}
\usepackage{amssymb}
\usepackage{amsfonts}

\usepackage[numbers]{natbib}

\usepackage{url}
\usepackage{hyperref}
\usepackage{booktabs}
\usepackage{multirow}
\newcommand{\nn}{\mathcal{N}}
\newcommand{\pp}{\mathcal{P}}
\newcommand{\ff}{\textsc{f}}

\makeatletter
\newcommand*{\rom}[1]{\expandafter\@slowromancap\romannumeral #1@}
\makeatother

\usepackage{adjustbox}

\usepackage[T1]{fontenc}
\usepackage[]{authblk}

\usepackage{abstract}

\begin{document}

\title{\textbf {Effect of minimal length uncertainty on the mass-radius relation of white dwarfs}}
\author{Arun Mathew%
\thanks{a.mathew@iitg.ernet.in} \ and Malay K. Nandy\thanks{mknandy@iitg.ernet.in}}
\affil{\normalsize Department of Physics, Indian Institute of Technology Guwahati, 781039 India.}
\date{October 10, 2017}    
\twocolumn[
  \begin{@twocolumnfalse}
    \maketitle
    \begin{abstract}
      Generalized uncertainty relation that carries the imprint of quantum gravity introduces a minimal length scale into the description of space-time. It effectively changes the invariant measure of the phase space through a factor $(1+\beta \mathbf{p}^2)^{-3}$ so that the equation of state for an electron gas undergoes a significant modification from the ideal case. It has been shown in the literature (Rashidi 2016) that the ideal Chandrasekhar limit ceases to exist when the modified equation of state due to the generalized uncertainty is taken into account. To assess the situation in a more complete fashion, we analyze in detail the mass-radius relation of Newtonian white dwarfs whose hydrostatic equilibria are governed by the equation of state of the degenerate relativistic electron gas subjected to the generalized uncertainty principle. As the constraint of minimal length imposes a severe restriction on the availability of high momentum states, it is speculated that the central Fermi momentum cannot have values arbitrarily higher than $p_{\rm max}\sim\beta^{-1/2}$. When this restriction is imposed, it is found that the system approaches limiting mass values higher than the Chandrasekhar mass upon decreasing  the parameter $\beta$ to a value given by a legitimate upper bound. Instead, when the more realistic restriction due to inverse $\beta$-decay is considered, it is found that the mass and radius approach the values close to $1.45$ M$_{\odot}$ and $600$ km near the legitimate upper bound for the parameter $\beta$. On the other hand, when $\beta$ is decreased sufficiently from the legitimate upper bound, the mass and radius are found to be approximately $1.46$ M$_{\odot}$ and $650$ km near the neutronization threshold.  
    \end{abstract}
    \vspace{0.5cm}
  \end{@twocolumnfalse}    
]

\saythanks  
\section{Introduction}
In recent years we find in the literature a contrasting perspective on the Heisenberg principle of uncertainty. Quantum theories of gravity, namely, string theory \citep{Amati1989,Konishi1990}, black hole physics \citep{Maggiore1993,Maggiore1994,Scardigli1999}, path-integral quantum gravity \citep{Padmanabhan1985a,Padmanabhan1985b,Padmanabhan1986,Padmanabhan1987}, and lattice quantum gravity \citep{Greensite1991}, predict the existence of a fundamental observable length in contrast with the Heisenberg uncertainty principle (HUP). Long ago, the existence of a fundamental length was shown by Mead \cite{Mead1964} by considering the uncertainty in measuring the position of a particle when it is observed through a Heisenberg microscope by a photon with which the particle is assumed to interact  both electromagnetically and gravitationally. It appears that the existence of a fundamental length scale is a model independent feature of quantum gravity and Garay \cite{Garay1995} has reviewed various routs in quantum gravity theories through which a minimal length uncertainty can be established. More recently, Adler \cite{Adler1999} obtained the generalised uncertainty relation following these arguments. 

This departure from the standard Heisenberg uncertainty principle suggests the breakdown of the notion of continuousness of space at length scales of the order of the Planck length $\ell_{P} = \sqrt{G \hbar/c^3}\approx1.6\times10^{-35}$ m. Among the proposals made to incorporate the concept of minimum length into quantum mechanics, Kempf \cite{Kempf1995} developed a theoretical framework admitting non-zero minimum uncertainties both in position and in momentum. In addition,  Pedram \cite{pedram2012a, pedram2012b} suggested a possible variant of deformed commutation relation that admits a non-zero minimum uncertainty in position and a maximal observable momentum. Assuming that these approaches consider the effects of quantum gravity correctly, we expect an observable deviation from the contemporary physics at ultra-fine length scales or ultra-high energies. Thus the best systems to search for such an anomaly are the ultra-dense compact stars that have offered themselves as model systems to investigate upon the fundamental principles of physics. Bertolami and Zarro \cite{Bertolami2010} demonstrated that the quantum gravity correction to a main sequence star is very small. However they argued that the case of white dwarfs would be peculiar since the Chandrasekhar limiting mass \cite{Chandrasekhar1939} is obtained by taking the limit of infinite density. 

Generalized uncertainty relation admitting a minimum observable length without any bound on the momentum was extensively studied by Maggiore \cite{Maggiore1993}, Kempf et al. \cite{Kempf1995} and  Hossenfelder \cite{Hossenfelder2006}. Utilising the commutation relations $[\hat{x}_i,\hat{p}_j]=i\hbar\delta_{ij}(1+\beta \hat{\textbf{p}}^2)$, $[\hat{p}_i,\hat{p}_j]=0$, and the non-commutative algebra $[\hat{x}_i,\hat{x}_j]=2i\hbar\beta(\hat{p}_i\hat{x}_j-\hat{p}_j\hat{x}_i)$, Chang et al. \cite{Chang2002} considered a classical analogue (Liouville theorem) of the time evolution and demanded invariance of the phase volume under infinitesimal time-translation. The invariant measure of the phase volume thus turned out to be $(1+\beta \textbf{p}^2)^{-3} d^3x \ d^3p$. 

 The magnitude of the quantum gravity parameter $\beta$  in the generalized uncertainty principle (GUP) is presently unknown and there exist estimates for upper (and lower) bounds. Brau \cite{Brau1999} calculated a correction to the Hydrogen atom energy levels due to GUP and compared the corresponding change in the 1S-2S transition with the experimental uncertainty in the frequency of the transition, yielding an uncertainty of $\Delta x_0\leq0.01$ fm. Subsequently, Brau and Buisseret \cite{brau2006} compared the first-order perturbation due to GUP in the energy spectrum of a neutron in a gravitational quantum well with the error bars in the energy spectrum obtained in the GRANIT experiment \citep{Nesvizhevsky2005} and obtained the upper bound for $\hbar^2\beta$ as $2\times10^{-5}$ fm$^2$. These estimates correspond to an upper bound for $\beta_0$ ($= \beta M_{P}^2c^2$) to be $7.6564\times10^{34}$. 
 
Das and Vagenas \cite{Das2008} obtained a few upper bounds on the GUP parameter $\beta_0$ by considering simple quantum mechanical systems. They showed that the accuracy of precision measurements of the Lamb shift in hydrogen atom predicted the upper bound $\beta_0<10^{36}$, which is compatible with the bound $\beta_0\leq10^{34}$ set by the electroweak length scale. On the other hand, the accuracy in the measurements of Landu levels using an STM suggested a higher value for the upper bound, namely, $\beta_0<10^{50}$. However, it may be noted that the accuracy in an STM is about $1$ part in $10^3$, and therefore this higher value of the bound is not as reliable as in the case of Lamb shift measurements where the accuracy is about $1$ part in $10^{12}$. In fact, they showed that if the GUP induced excess current in the tunnelling of electrons in an STM cannot be measured in a time-scale of one year, the upper bound shifts to $\beta_0<10^{21}$, whereas if this time-scale is chosen to be $1$ second, the upper bound is $\beta_0<10^{29}$. There is also an estimate, namely $\beta_0= 82\pi/5$, from a calculation based on black hole thermodynamics \citep{Scardigli2017}. 

On the basis of Hagedron temperature for relativistic strings, Wang et al. \cite{Wang2010} suggested a lower bound for $\beta_0$, namely, $\beta_0>10^4$. In addition,  they considered the case of ultra-relativistic ($E=pc$) Fermi gas at $T=0$ and, approximating the density to be constant, obtained corrections in the leading order of the parameter $\beta$. They found that the GUP correction resists gravitational collapse as it tends to raise the mass of white dwarfs by a small amount, namely, $\Delta M\sim 10^{-10}M_{\odot}$ for $\beta_0=10^{36}$. They further showed that the radius tends to diverge for low and high values of $\beta_0$ whereas a minimum radius $\beta_0 \ell_P$ occurs in between implying the absence of a singularity (zero radius), which is of course a consequence of the minimum length uncertainty. 

Ali \cite{Ali2011}, on the other hand, used a different uncertainty relation that corresponds to a commutative space and obtained an invariant measure of the phase volume, namely $(1-\alpha p)^{-4}d^3x \ d^3p$, along the lines of Chang et al. \cite{Chang2002}. In a successive work, Ali and Tawfik \cite{Ali2013} considered the case of white dwarfs assuming uniform density with ultra-relativistic Fermi gas. In a similar manner to Wang et al. \cite{Wang2010}, they calculated the mass of the white dwarf and found the GUP correction to be higher, namely $\Delta M\sim 10^{-5}M_{\odot}$ for $\alpha M_P c=10^{17}$, whereas the radius decreases with increasing Fermi energy.  

There is an alternative approach through which a high energy scale (or a small length scale) can enter into the description. This constitutes the postulate of a non-commutative space-time breaking Lorentz invariance and leading to a deformed dispersion relation (between energy and momentum) involving the high energy scale. Following this approach, Camacho \cite{Camacho2006} obtained the mass-radius relation of white dwarfs assuming their densities to be uniform. For a negative definite deformation parameter occurring in the dispersion relation, he predicted the existence of white dwarfs with very large radii and masses close to the Chandrasekhar mass. Gregg and Major \cite{Gregg2009} considered the realistic case of white dwarfs of non-uniform density and reported results that were quite distinct from the previous predictions. They arrived at new limiting masses about 10\% higher (lower) than the Chandrasekhar mass limit for positive (negative) deformation parameter occurring in the dispersion relation. A different dispersion relation admitting an upper bound for the momentum was analyzed by Bertolami and Zarro \cite{Bertolami2010}. Their investigation suggested that the non-commutative contribution brings in added stability in white dwarfs. Amelino-Camelia et al. \cite{Camelia2012} analyzed the problem of employing both the deformed dispersion relation and deformed momentum space measure to obtain the mass-radius relation. This method introduced  a shift in the deformation parameter that did not bring in any qualitative change in the mass-radius relationship. 

Rashidi \cite{Rashidi2016}, on the other hand, considered the case of Helium white dwarfs assuming an ultra-relativistic ($E=pc$) Fermi gas and, using the generalized uncertainty invariant measure $(1+\beta p^2)^{-3}d^3x\ d^3p$, obtained the equation of state in the ideal degenerate case. He further considered the hydrostatic equilibrium with a non-uniform density and analyzed a generalized version of the Lane-Emden equation as an initial value problem. In the limit of infinite central Fermi momentum, the differential equation reduces to the Lane-Emden equation of index zero, immediately leading to the mass and radius behaving as $M\sim p_{\ff c}^{3/2}$ and $R\sim p_{\ff c}^{1/2}$. It follows that both mass and radius of  white dwarfs increase unboundedly with an indefinite increase in central Fermi momentum $p_{\ff c}$ so that the ideal Chandrasekhar limit ceases to exit. 

In this paper, we shall examine in detail the effect of generalised uncertainty on Helium white dwarfs with invariant measure  $(1+\beta p^2)^{-3}d^3x\ d^3p$. We shall take the full equation of state for a completely degenerate electron gas with the relativistic dispersion relation ($E^2 = p^2 c^2 + m_e^2 c^4$) and study the implication of the quantum gravity parameter $\beta$ (or $\beta_0= \beta M_{P}^2c^2$) on the mass-radius relation. We first perform a heuristic analysis by taking the density to be uniform and consider the effect of increasing the Fermi momentum $p_{\ff}$ unboundedly. We find that the mass and radius behave as $M\sim p_{\ff}^{3/2}$ and $R\sim p_{\ff}^{1/2}$. To analyse the problem in somewhat more detail, we approximately solve the equation of hydrostatic equilibrium relaxing the previous assumption of uniform density.  We find that, in the asymptotic limit of high central Fermi momentum, the white dwarf has a core of approximately uniform density and an envelope where the density falls off. It is interesting to note that the mass and radius of the core behave as $\sim p_{\ff c}^{3/2}$ and $\sim p_{\ff c}^{1/2}$. Thus, in both of these analyses, the mass and radius approach the same asymptotic behaviour as obtained by Rashidi \cite{Rashidi2016}.

In the next program of our calculations, we take the realistic case of varying density and use the exact equations of hydrostatic equilibrium. We first solve these equations numerically and determine the mass-radius relation for different values of the GUP parameter $\beta_0$ to assess the situation of the mass-radius relationship. It is found that, with increase in the central Fermi momentum, the radius approaches a minimum value while the mass increases slowly. However, beyond this minimum radius value, the mass and radius start to increase unboundedly. This latter regime is identified with the previous asymptotic analysis for infinitely large Fermi momentum.

We next note that the GUP factor $(1+\beta \mathbf{p}^2)^{-3}$ puts a severe restriction on the availability of ultra-high momentum states as this factor reduces to a small value for $\sqrt{\beta}p=1$. We thus consider four different cases with the maximum value $(\sqrt{\beta}p)_{\rm max}$ =$1$, $1.25$, $1.5$, and $3$. In each of these cases we calculate the mass and radius for different choices of $\beta_0$. Within these restrictions we find that both mass and radius are finite if the upper bound of $\beta_0$ is taken to be $10^{36}$. However, if $(\sqrt{\beta}p)_{\rm max}$ is increased unboundedly, the mass and radius values also increase unboundedly.  

In a white dwarf, a realistic upper bound for the central Fermi energy is the threshold energy for inverse $\beta$-decay. We therefore next take the central Fermi momentum to be the neutronization threshold  given by $E_{\rm N}^2=p_N^2c^2+m_e^2 c^4$. Consequently we take different cases corresponding to  $\beta_0$ ranging from $10^{44}$ to $10^{36}$. We find that both mass and radius approach approximately constant values, namely, $1.45$ $M_{\odot}$ and  $600$ km, as the GUP parameter $\beta_0$ is decreased below $10^{38}$. However, if $\beta_0$ is increased unboundedly, both the mass and radius values also increase unboundedly.  

The rest of the paper is organized as follows. In Section 2, we find the expression for density and pressure in terms of the Fermi momentum for an electron gas with the GUP invariant measure, where we also analyze the asymptotic behavior of these quantities and discuss the restriction on momentum states by the GUP. In Section 3, we consider the equation of hydrostatic equilibrium for a white dwarf with Newtonian gravity and work out a few approximate  solutions for the mass and radius, where we also report an exact solution by a numerical scheme. In Section 4, we numerically solve the equation of hydrostatic equilibrium with restrictions due to GUP and neutronization. Finally, discussion and conclusion are given in Section 5. 

\section{Fermi gas GUP equation of state} 
In this section, we derive the expression for the number density $n$ of electrons and its pressure $P$ in a degenerate electron gas using the grand canonical partition function. The effect of GUP commutation relation is taken into account by considering the modified invariant measure. The asymptotic behavior for $n$ and $P$ are analyzed in Section 2.2. The GUP modification of density of states puts a severe restriction on the high momentum states which is discussed in the Section 2.3, where we also lay out the assumption on the maximum momentum.
\subsection{Grand canonical ensemble}
The Grand partition function \citep{Landau1969} for a Fermi gas at temperature $T$ can be written as 
\begin{equation}\label{eq:gpf}
\mathcal{Z} = \prod_{\mathbf{p}} \left[\sum_{n_{\mathbf{p}}} \exp \left\{-\frac{(E_\mathbf{p}-\mu)n_\mathbf{p}}{k_B T}\right\} \right]^g
\end{equation}
where $E_{\mathbf{p}}=\sqrt{\mathbf{p}^2c^2+m_e^2 c^4}$, $n_{\mathbf{p}}$ is the number of fermions in the momentum state $|\mathbf{p}\rangle$ with degeneracy $g$, $\mu$ is the chemical potential, and $k_B$ is the Boltzmann constant. Since $n_{\mathbf{p}}=0,1$ and $g=2$ for electrons, the grand potential $\Omega=-k_B T \ln \mathcal{Z}$ turns out to be 
\begin{equation}\label{eq:gpoten}
\Omega = -2k_B T\sum_{\mathbf{p}} \ln \left[1 + \exp\left\{-\frac{(E_{\mathbf{p}}-\mu)}{k_B T}\right\}\right].
\end{equation}
so that the number of electrons is given by 
\begin{equation}\label{eq:number}
N = -\frac{\partial \Omega}{\partial \mu} =2 \sum_{\mathbf{p}}\frac{1}{\exp\left( \frac{E_\mathbf{p}-\mu}{k_B T}\right)+1}.
\end{equation}

Since  the GUP formalism demands that the number of states available be transformed according to 
\begin{equation}\label{eq:integral}
\sum_{\mathbf{p}} \equiv \int \frac{V d^3p}{h^3} \frac{1}{(1+\beta \mathbf{p}^2)^3}
\end{equation}
where $h$ is the Planck's constant, we therefore obtain the number density $n=N/V$ as 
\begin{equation}\label{eq:n1}
n = \frac{2}{h^3} \int_0^\infty \frac{4\pi p^2 dp}{(1+\beta p^2)^3} \frac{1}{\exp\left( \frac{E_{\mathbf{p}}-\mu}{k_B T}\right)+1}.
\end{equation}

\begin{figure}
\centering
\includegraphics[width=8cm]{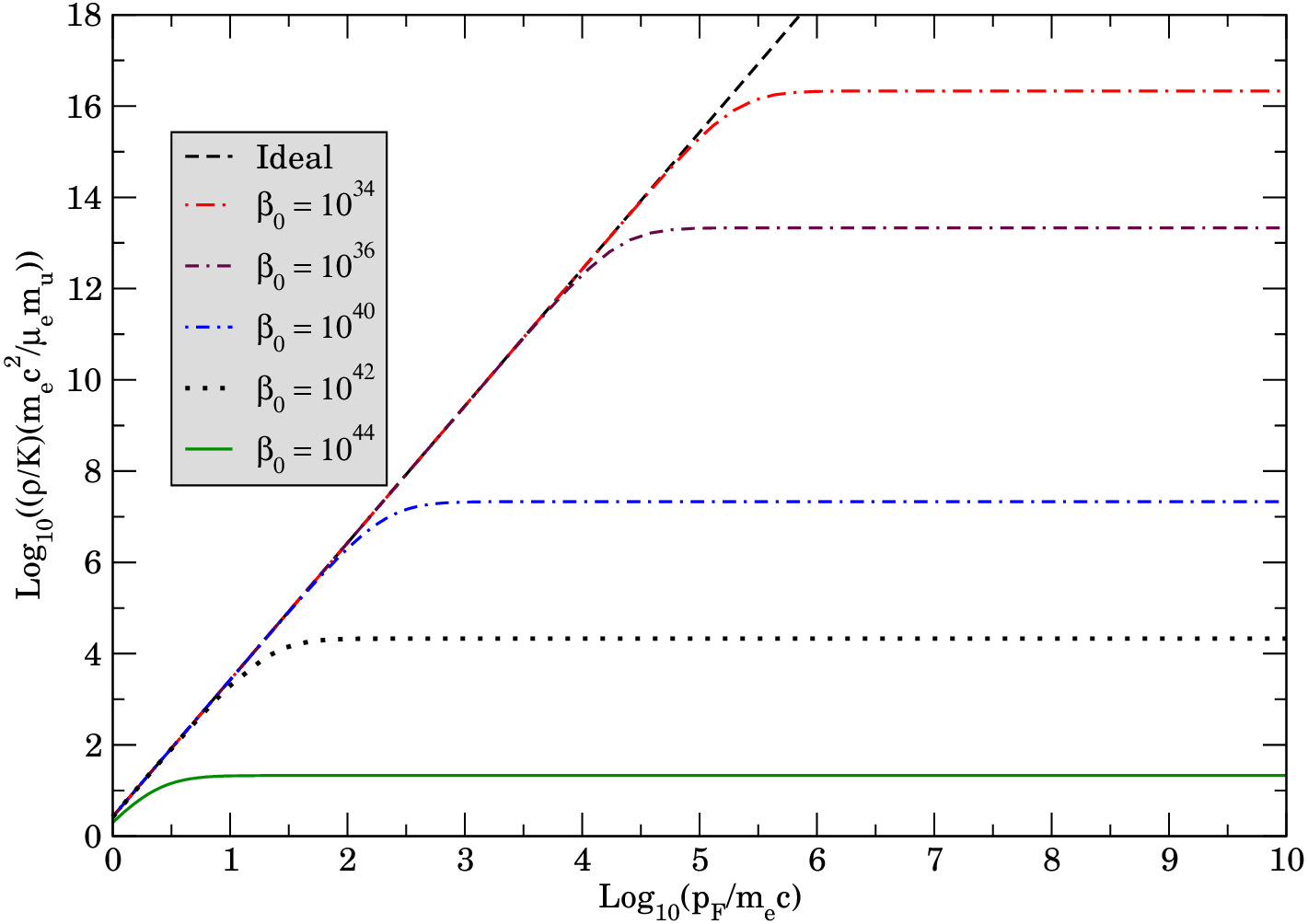}
\caption{\small  Dimensionless density $(\rho/K)(m_ec^2/\mu_e m_u)$ versus dimensionless Fermi momentum $p_\ff/m_e c$ for various values of $\beta_0$.}
\label{figure:1}
\end{figure}

Since $\Omega=-PV$, we readily obtain the pressure as 
\begin{equation}\label{eq:P1}
P = \frac{2k_B T}{h^3} \int_0^\infty \frac{4\pi p^2 dp}{(1+\beta p^2)^3} \ln \left[1 + \exp\left\{-\frac{(E_{\mathbf{p}}-\mu)}{k_B T}\right\}\right]
\end{equation}

As we can treat the electron gas in white dwarfs to be completely degenerate, we take the limit $T\rightarrow0$ in the above integrands and obtain
\begin{equation}\label{eq:n2}
n=\frac{8\pi}{h^3}\int_0^{p_{\ff}} \frac{p^2dp}{(1+\beta p^2)^3}
\end{equation}
and 
\begin{equation}\label{eq:P2}
P = \frac{8\pi}{h^3} \int_0^{p_\ff} \frac{p^2dp}{(1+\beta p^2)^3} \left(E_\ff- E_{\mathbf{p}}\right),
\end{equation}
where $E_\ff=\sqrt{p_\ff^2c^2+m^2 c^4}$ is the Fermi energy (or the chemical potential $\mu$ at $T=0$), and $p_\ff$ is the Fermi momentum. 

Integrating the above two equations, Eqs~(\ref{eq:n2}) and (\ref{eq:P2}), we obtain 
\begin{equation}\label{eq:npF}
n(p_\ff)=\frac{\pi}{(h\sqrt{\beta})^3} \left\{ \tan^{-1} (\sqrt{\beta} p_\ff)-\frac{(1- \beta p_\ff^2)}{(1+\beta p_\ff^2)^2}\right\}
\end{equation}
and 
\begin{align}\label{eq:PpF}
&P(p_\ff)= \frac{\pi}{(h\sqrt{\beta})^3}\sqrt{p_\ff^2 c^2+m_e^2 c^4} \left\{\tan^{-1} (\sqrt{\beta}p_\ff) \right. \nonumber \\
&\left.  -\frac{\sqrt{\beta}p_\ff}{(1-\beta m_e^2 c^2)(1+\beta p_\ff^2)}\right\} +\frac{\tanh^{-1} \left(\frac{p_\ff \sqrt{1-\beta m_e^2 c^2}}{\sqrt{p_\ff^2+m_e^2 c^2}}\right)}{(1-\beta m_e^2 c^2)^{\frac{3}{2}}}. 
\end{align}

Using the dimensionless quantities $\xi=p_\ff/m_e c$ and $\alpha=m_e c\sqrt{\beta}$, we can express  the above equations as
\begin{equation}\label{eq:n}
 n(\xi)=\frac{K}{m_e c^2}\nn(\xi)
\end{equation}
and 
\begin{equation}\label{eq:P}
P(\xi) = K \pp(\xi)
\end{equation}
with $K=\pi m_e^4 c^5/h^3=1.8007\times10^{22}$ N/m$^2$ having the dimension of pressure and 
\begin{equation}\label{eq:curlyN}
 \nn(\xi)=\frac{1}{\alpha^3}\left\{ \tan^{-1} (\alpha \xi)-\frac{\alpha \xi(1- \alpha^2 \xi^2)}{(1+\alpha^2 \xi^2)^2}\right\},
\end{equation}
\begin{align}\label{eq:curlyP}
\pp(\xi) = &\frac{\sqrt{1+\xi^2}}{\alpha^3} \left\{\tan^{-1} (\alpha \xi)  -\frac{\alpha \xi}{(1-\alpha^2)(1+\alpha^2 \xi^2)}  \right\} \nonumber \\ 
&+\frac{\tanh^{-1} \left(\frac{\xi \sqrt{1-\alpha^2 }}{\sqrt{1+\xi^2}}\right)}{(1-\alpha^2)^{\frac{3}{2}}} .
\end{align}
Knowing the electron number density $n$, the mass density $\rho$ can be written as $\rho=\mu_e m_u n$. Consequently 
\begin{equation}\label{eq:rho}
\rho(\xi) = K \left(\frac{\mu_e m_u}{m_e c^2}\right) \nn(\xi)
\end{equation}
with $\mu_e=A/Z$ representing the number of nucleons per electron and $m_u=1.6605\times10^{-27}$ kg is the atomic mass unit. 

\begin{figure}
\centering
\includegraphics[width=8cm]{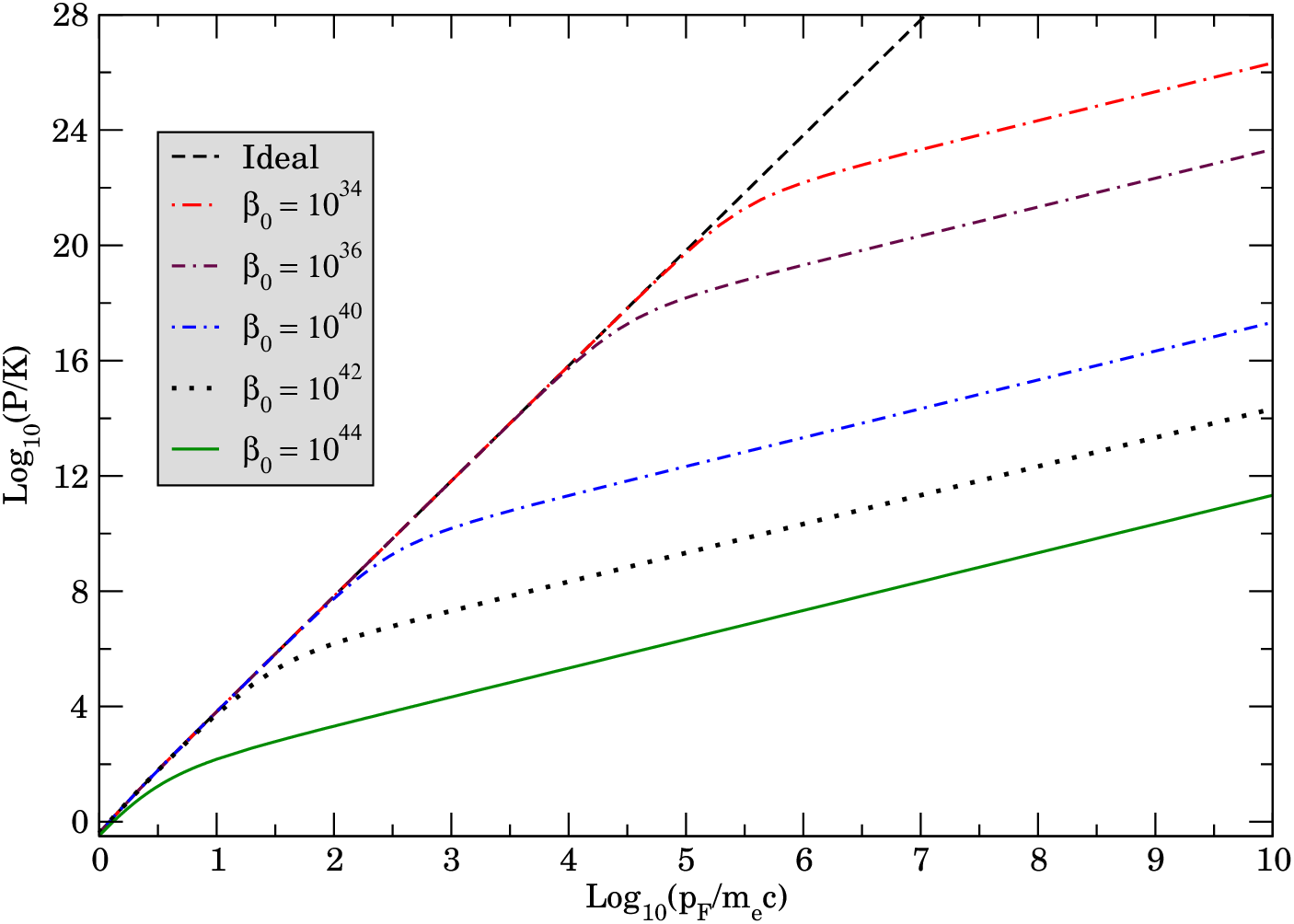}
\caption{\small Dimensionless pressure $P/K$ versus dimensionless Fermi momentum $p_\ff/m_e c$ for various values of $\beta_0$.}
\label{figure:2}
\end{figure}

The quantum gravity parameter $\sqrt{\beta}  = \sqrt{\beta_0}/M_{P}c$ is determined by the dimensionless parameter $\beta_0$ and the Planck mass $M_{P}=\sqrt{\hbar c/G}=2.1765\times10^{-8}$ kg. We shall choose different values of $\beta_0$ to tune the parameter $\beta$. Table \ref{table:1} shows different values of $\alpha$ corresponding to $\beta_0$. Figure \ref{figure:1} shows how the density $\rho$ changes with respect to $\xi=p_\ff/m_e c$ for different values of the parameter $\beta_0$. It is noticed that the density saturates to different constant values for large values of $\xi$ corresponding to different choices for $\beta_0$. As $\beta_0$ is decreased, the density resembles more closely to the $\beta_0=0$ case and the (higher) saturation value of $\rho$ shifts to higher values of $\xi$. Figure \ref{figure:2} displays the variation of pressure $P$ with respect to $\xi=p_\ff/m_e c$ for different value of $\beta_0$. It is seen that, for higher values of $\xi$, pressure varies proportionally to $\xi$. The $\beta_0=0$ line corresponds to $P\sim\xi^4$ which is true for higher values of $\xi$. We see that, as $\beta_0$ is decreased, the linear regime shifts to higher and higher values of $\xi$. 

\begin{table}
\centering
\caption{\small Different values of $\beta_0$ and the corresponding $\alpha$ values with asymptotic values of $\nn(\xi)$ and $\pp(\xi)$ for large values of $\xi$.}
\label{table:1}
\vspace{0.1cm}
\begin{tabular}{cccc}
\hline
\hline
$\beta_0$  & $\frac{\alpha}{4.1854}$     & $\frac{\nn(\infty)}{2.1425}$ & $\pp(\xi)=\nn(\infty)\xi+B(\alpha)$ \\[0.5ex] 
\hline
$10^{44}$   & $10^{-1}$    & $10^{01}$      & $2.1425\times10^{01}\xi-7.0071\times10^{01}$  \\    
$10^{42}$   & $10^{-2}$    & $10^{04}$      & $2.1425\times10^{04}\xi-6.5232\times10^{05}$  \\   
$10^{40}$   & $10^{-3}$    & $10^{07}$      & $2.1425\times10^{07}\xi-6.5176\times10^{09}$  \\ 
$10^{38}$   & $10^{-4}$    & $10^{10}$     & $2.1425\times10^{10}\xi-6.5176\times10^{13}$ \\
$10^{36}$   & $10^{-5}$    & $10^{13}$     & $2.1425\times10^{13}\xi-6.5176\times10^{17}$ \\
\hline
\hline
\end{tabular}
\end{table}

\subsection{Asymptotic behavior}

The above expression for number density approaches a constant value $n(\infty)=\pi^2 m_e^3 c^3/2h^3\alpha^3$ as the function $\nn(\xi)$ approaches a constant 
\begin{equation}\label{eq:Ninfinity}
 \nn(\infty)=\frac{\pi}{2\alpha^3}
 \end{equation} 
in the limit $\xi\rightarrow\infty$. This situation is unlike the HUP electron number density $n_{\textsc{hup}}(\xi) = 8\pi m_e^3 c^3\xi^3/3h^3$ that approaches infinity in the same limit. The saturation values $\rho(\infty)$ can be clearly seen in Figure \ref{figure:1} for different value of $\beta_0$.

Minimum distance between the particles is expected to be $d_{\rm min}=(16\pi)^{1/3}\Delta x_0$ where $\Delta x_0$ is the minimum uncertainty in position imposed by the GUP. This minimum distance will be approached in an electron gas of extremely high density which will be the case when the Fermi momentum is extremely high . This is consistent with the fact that the GUP number density  given by Eq.~(\ref{eq:n}) approaches a constant value $n(\infty)$ in the limit $\xi\rightarrow\infty$. 

The original degenerate case is expected to be reproduced in the limit $\alpha\rightarrow0$, corresponding to low values of $\beta_9$. The leading order term in the expansion of the function $\nn(\xi)$ for finite values of $\xi$ gives $\nn(\xi)\rightarrow3\xi^3/8$ yielding $n(\xi)\rightarrow n_{\textsc {hup}}(\xi) = 8\pi m_e^3 c^3\xi^3/3h^3$ which is the original degenerate case. Since $n(\xi)$ approaches the same limit in the limit  $\xi\rightarrow0$, the number density $n(\xi)$  approaches $n_{\textsc {hup}}(\xi)$ for low values of $\xi$. This behavior can be seen in the plot for $\rho(\xi)$ in Figure \ref{figure:1} for various values of $\beta_0$ in the region of low $\xi$. 

In a similar manner, the original degenerate pressure is recovered in the limit $\alpha\rightarrow0$, yielding $P(\xi)\rightarrow P_{\textsc{hup}}(\xi)$, namely
\begin{equation}\label{eq:PHUP}
 P_{\textsc{hup}}(\xi)=\frac{K}{3}\left\{\xi\sqrt{1+\xi^2}(2\xi^2 -3)+3\sinh^{-1}\xi\right\}
\end{equation}

For moderately large values of $\xi$, the HUP pressure behaves as $P_{\textsc{hup}}(\xi)\sim2K\xi^4/3$. However, for large values of $\xi$, the GUP pressure given by Eq.(\ref{eq:P}) goes asymptotically as  
\begin{equation}\label{eq:Pinfinity}
P(\xi)\rightarrow K\left\{\nn(\infty)\xi+B(\alpha)\right\}
\end{equation}
where
\begin{equation}\label{eq:B}
B(\alpha)=\frac{\tanh^{-1}(\sqrt{1-\alpha^2})}{(1-\alpha^2)^{3/2}}-\frac{1}{\alpha^4}\frac{(2-\alpha^2)}{(1-\alpha^2)}.
\end{equation}

This linear behaviour of the GUP pressure (with respect to $\xi$) can clearly be seen in Figure \ref{figure:2} where the curve departs from the $\xi^4$ regime to an $\xi$ regime changing its slope from $4$ to $1$. Thus, unlike the HUP pressure,  the GUP pressure increases more slowly for large values of $\xi$. It may also be noted that, in a linear plot for Eq.~(\ref{eq:Pinfinity}), the slope and intercept both will depend on the value of $\alpha$.

\subsection{GUP restriction on momentum}

\begin{figure}
\centering
\includegraphics[width=8cm]{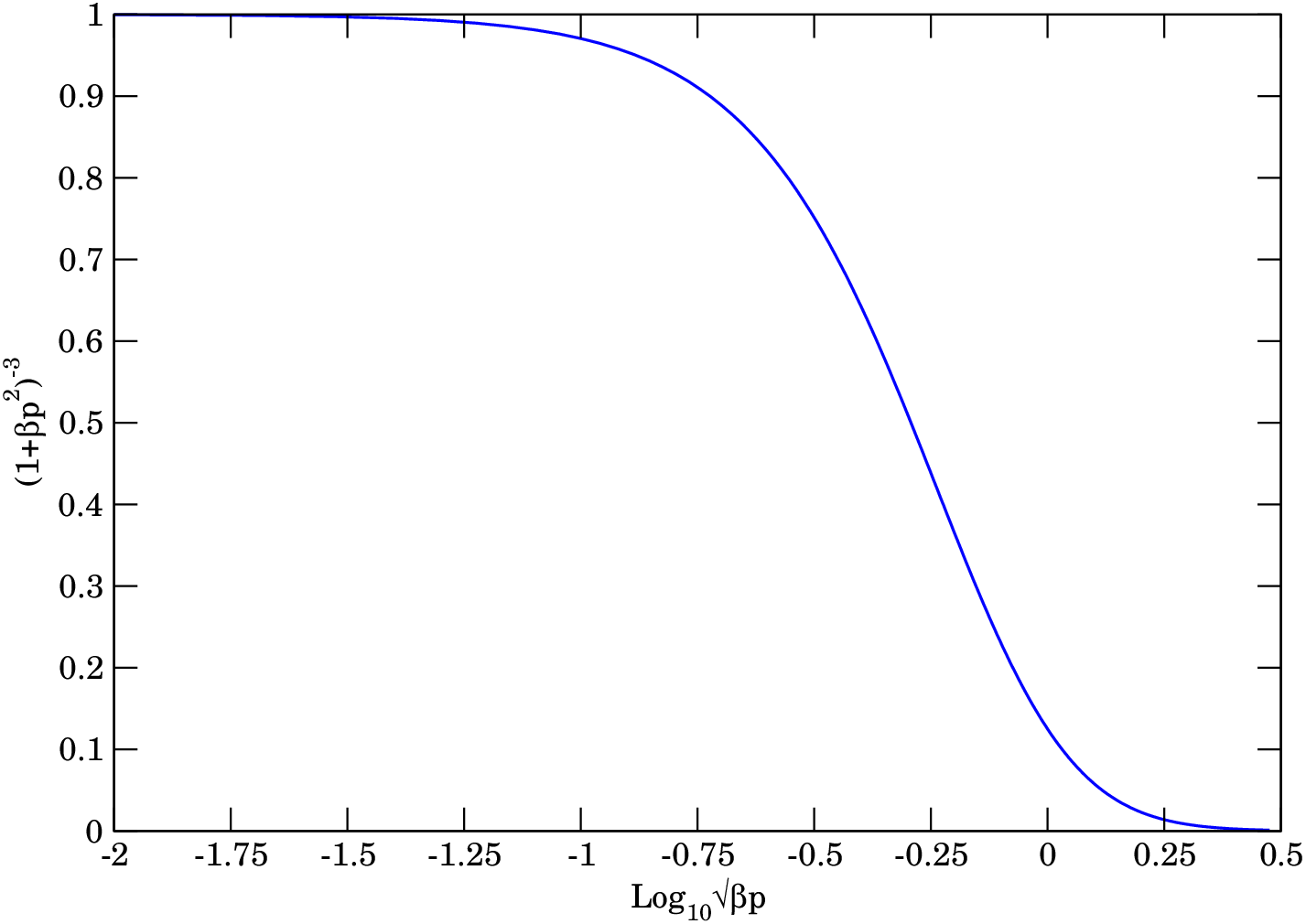}
\caption{\small Variation of the GUP deformation factor $f=(1+\beta p^2)^{-3}$ with respect to the dimensionless quantity $\sqrt{\beta}p$ showing suppression of  the high momentum states.}
\label{figure:3}
\end{figure}

The center of a white dwarf is expected to have a high value of Fermi momentum. However, the GUP deformation factor $f(p)=(1+\beta p^2)^{-3}$ imposes a severe restriction on the allowed values of momenta. Figure \ref{figure:3} shows the variation of $f(p)$ with respect to the dimensionless quantity $\sqrt{\beta}p$. It is seen that this factor suppresses the high momentum states as the curve decreases strongly for high values of momenta. For example, for $\sqrt{\beta}p=1$, this factor reduces to $f=0.125$; for $\sqrt{\beta}p=1.25$, which corresponds to  $\log_{10}\sqrt{\beta}p=0.097$, $f=0.059$; for $\sqrt{\beta}p=1.50$, or $\log_{10}\sqrt{\beta}p=0.176$, $f$ reduces to $0.029$, and for $\sqrt{\beta}p=3.00$, or  $\log_{10}\sqrt{\beta}p=0.477$, $f$ reduces to $0.001$. Consequently there is less or even negligible contribution from momentum states belonging to momentum values higher than these values. In fact, this behavior suggests that the integral in momentum space has an effective cutoff $p_{\rm max}\sim\beta^{-1/2}$ and thus there is negligible contribution from momenta higher than $p_{\rm max}$. It is thus important to take values of Fermi momentum $p_F$ equal to or lower than $p_{\rm max}$. We shall therefore choose the maximum Fermi momentum $p_F$ (at the center of white dwarf) corresponding to these four maximum values of $(\sqrt{\beta} p)_{\rm max}$. These four choices for the maximum values are shown in Table \ref{table:2} that also shows the maximum values $\xi_{\rm max}$, $\nn(\xi_{\rm max})$ and $\pp(\xi_{\rm max})$ corresponding to different choices of $\beta_0$. 

\begin{table}
\centering
 \caption{Four choices of the maximum value of  $\sqrt{\beta} p$  as suggested by Figure \ref{figure:3}  beyond which the momentum states have small contributions. Different choices of  $\beta_0$  correspond to different values of  $\xi_{\rm max}$.  The corresponding values of the dimensionless number density $\nn(\xi_{\rm max})$ and pressure $\pp(\xi_{\rm max})$ are also shown.}
 \label{table:2}
 \vspace{0.1cm}
\begin{adjustbox}{width=0.45\textwidth}
\begin{tabular}{ccccc}
\hline
\hline
$\sqrt{\beta}p_{\rm max}$&$\beta_0$ &$\xi_{\rm max}$&$\nn(\xi_{\rm max})$& $\pp(\xi_{\rm max})$ \\
\hline
                         & $10^{44}$ & $2.389\times10^0$    & $1.071\times10^{01}$         & $7.951\times10^{00}$\\    
     		         & $10^{42}$ & $2.389\times10^1$    & $1.071\times10^{02}$         & $9.280\times10^{04}$\\    
     1.00                & $10^{40}$ & $2.389\times10^2$    & $1.071\times10^{07}$         & $9.300\times10^{08}$\\
                         & $10^{36}$ & $2.389\times10^4$    & $1.071\times10^{13}$      & $9.300\times10^{16}$\\
                         & $10^{34}$ & $2.389\times10^5$    & $1.071\times10^{16}$      & $2.146\times10^{20}$\\
			 \hline    
			 & $10^{44}$ & $2.986\times10^0$    & $1.368\times10^{01}$       & $1.482\times10^{01}$\\    
                         & $10^{42}$ & $2.986\times10^1$    & $1.368\times10^{04}$       & $1.661\times10^{05}$\\    
      1.25               & $10^{40}$ & $2.986\times10^2$    & $1.368\times10^{07}$       & $1.663\times10^{09}$\\
                         & $10^{36}$ & $2.986\times10^4$    & $1.368\times10^{13}$    & $1.663\times10^{17}$\\
                         & $10^{34}$ & $2.389\times10^5$    & $1.363\times10^{16}$    & $1.663\times10^{21}$\\
                         \hline
			 & $10^{44}$ & $3.584\times10^0$     & $1.583\times10^{01}$        & $2.329\times10^{01}$  \\    
                         & $10^{42}$ & $3.584\times10^1$     & $1.583\times10^{04}$        & $2.545\times10^{05}$ \\    
     1.50                & $10^{40}$ & $3.584\times10^2$     & $1.583\times10^{07}$        & $2.548\times10^{09}$\\
                         & $10^{36}$ & $3.584\times10^4$     & $1.583\times10^{13}$     & $2.548\times10^{17}$\\
                         & $10^{34}$ & $3.584\times10^5$     & $1.583\times10^{16}$     & $2.548\times10^{21}$\\
			\hline
			 & $10^{44}$ & $7.168\times10^0$    & $2.031\times10^{01}$         & $8.935\times10^{01}$\\    
                         & $10^{42}$ & $7.168\times10^1$    & $2.031\times10^{04}$         & $9.274\times10^{05}$\\    
     3.00                & $10^{40}$ & $7.168\times10^2$    & $2.031\times10^{07}$         & $9.278\times10^{09}$\\
                         & $10^{36}$ & $7.168\times10^4$    & $2.031\times10^{13}$      & $9.278\times10^{17}$\\
                         & $10^{34}$ & $7.168\times10^5$    & $2.031\times10^{16}$      & $9.278\times10^{21}$\\
\hline
\hline
\end{tabular}
\end{adjustbox}
\end{table}

\section{White Dwarfs with GUP}
In this section we obtain the mass-radius relation of Helium white dwarfs in the Newtonian gravity employing the modified equation of state following from the generalized uncertainty principle as obtained in the previous section. The condition of hydrostatic equilibrium for a spherical distribution of matter is given by 
\begin{equation}\label{eq:HS1}
\frac{dP}{dr}=-\frac{G m(r)\rho(r)}{r^2}
\end{equation}
with
\begin{equation}\label{eq:HS2}
\frac{dm}{dr}=4\pi\rho(r)r^2.
\end{equation}
The above two equations can be combined to obtain
\begin{equation}\label{eq:CHS}
\frac{1}{r^2}\frac{d}{dr}\left(\frac{r^2}{\rho}\frac{dP}{dr}\right)+4\pi G \rho = 0
\end{equation}
Substituting Eqs.~(\ref{eq:rho}) and (\ref{eq:P}) in Eq.~(\ref{eq:CHS}) yields 
\begin{equation}\label{eq:CHSS}
\frac{1}{4\pi G K}\left(\frac{m_e c^2}{\mu_e m_u}\right)^2\left[\frac{1}{r^2}\frac{d}{dr}\left(\frac{r^2\pp'(\xi)}{\nn(\xi)}\frac{d\xi}{dr}\right)+\nn(\xi)\right] = 0
\end{equation}
where $\pp'(\xi)=d\pp/d\xi$. Using the dimensionless variable  $x=r/R_0$, where $R_0=\frac{1}{\sqrt{4\pi G K}}\left(\frac{m_e c^2}{\mu_e m_u}\right)$, the above equation reduces to
\begin{equation}\label{eq:CHSD}
\frac{1}{x^2}\frac{d}{dx}\left(\frac{x^2\pp'(\xi)}{\nn(\xi)}\frac{d\xi}{dx}\right)+\nn(\xi)= 0.
\end{equation}

From Eqs.~(\ref{eq:curlyP}) and (\ref{eq:curlyN}), we find $\pp'(\xi)=\frac{\xi}{\sqrt{1+\xi^2}}\nn(\xi)$, so that the above equation reduces to 
\begin{equation}\label{eq:CHSR}
\frac{1}{x^2}\frac{d}{dx}\left(x^2\frac{d\sqrt{1+\xi^2}}{dx}\right)+\nn(\xi)= 0.
\end{equation}

Before considering the complete solution, we shall first analyze the problem heuristically and then consider the asymptotic behaviour of the above equation in the extreme limits $\xi\rightarrow0$ and $\xi\rightarrow\infty$.

\subsection{Heuristic treatment} 

We first analyze the problem heuristically by assuming the density to be uniform (that is, a uniform value of Fermi momentum) and obtain an approximate mass-radius relation as follows.  Using $n=\frac{K}{m_e c^2} \nn(\xi)$ and $n=\frac{3}{4\pi}\frac{m_ec^2}{\mu_e m_u}\frac{M}{R^3}$, and expanding $\nn(\xi)$ as $\nn(\xi)=\frac{\pi}{2\alpha^3}+\frac{8}{3\alpha^6}\frac{1}{\xi^3}-\ldots$ for large values of $\xi$, we obtain

\begin{equation}\label{eq:xi}
\xi^3=\frac{16}{3\pi\alpha^3} \frac{1}{ \left(1-\frac{m_ec^2}{\mu_e m_u}\frac{3}{4\pi K}\frac{M}{R^3}\right)}=\left(\frac{16}{3\pi\alpha^3}\right) \frac{1}{ \left(1-\frac{6\alpha^3}{\pi}\frac{\tilde{M}}{\tilde{R}^3}\right)}
\end{equation}
where $\tilde{M}=M/M_0$ and $\tilde{R}=R/R_0$ with $M_0=\frac{1}{\sqrt{4\pi K}}\left(\frac{1}{G}\right)^{3/2}\left(\frac{m_e c^2}{\mu_e m_u}\right)^2$.

The pressure can be estimated from equating the work done in compressing the star from infinite dilution to a radius $R$ in the absence of gravity and by equating it with the gravitational self-energy, yielding \citep{Huang1987}
\begin{equation}\label{eq:HSfs}
P=\frac{\gamma}{4\pi}\frac{GM_{\rm}^2}{R_{\rm}^4} = K \gamma \frac{\tilde{M}^2}{\tilde{R}^4}
\end{equation}
where $\gamma$ is a constant of $\mathcal{O}(1)$. Expanding $\pp(\xi)$ in the limit of large $\xi$, we have 
\begin{equation}\label{eq:curlyPexp}
\pp=\frac{\pi}{2\alpha^3}\xi + B(\alpha) + \frac{\pi}{4\alpha^3}\frac{1}{\xi}+\ldots
\end{equation}
Eliminating $\xi$ from Eqs. (\ref{eq:xi}) and (\ref{eq:curlyPexp}), and using (\ref{eq:HSfs}), we obtain
\begin{equation}\label{eq:HSfsS}
\gamma \frac{\tilde{M}^2}{\tilde{R}^4}=\frac{C(\alpha)}{ \left(1-\frac{6\alpha^3}{\pi}\frac{\tilde{M}}{\tilde{R}^3}\right)^{1/3}} + B(\alpha) + A(\alpha) \left(1-\frac{6\alpha^3}{\pi}\frac{\tilde{M}}{\tilde{R}^3}\right)^{1/3}
\end{equation}
where $A(\alpha)=\frac{\pi}{4\alpha^2}\left(\frac{3\pi}{16}\right)^{1/3}$ and $C(\alpha)=\frac{\pi}{2\alpha^4}\left(\frac{16}{3\pi}\right)^{1/3}$.
A solution to the above equation can be obtained in parametric form by defining $\left(1-\frac{6\alpha^3}{\pi}\frac{\tilde{M}}{\tilde{R}^3}\right)^{1/3}=z$ and $\frac{\tilde{M}^2}{\tilde{R}^4}=\frac{y}{z}$ so that it can be written in the form 
\begin{equation}\label{eq:yz}
\gamma y = A(\alpha)z^2 + B(\alpha)z+C(\alpha)
\end{equation}
 The $z^2$ term is negligible compared to the other terms for $z<1$. For example, for $\alpha=0.4185$, $A(\alpha)=3.7590$, $B(\alpha)=-70.0968$ and $C(\alpha)=61.0878$. Thus $y$ is related to $z$ almost linearly with a negative slope. 
 
Since $\nn(\xi)<\frac{\pi}{2\alpha^3}$, it follows that $\frac{6\alpha^3}{\pi}\frac{\tilde{M}}{\tilde{R}^3}<1$, so that $z>0$. Moreover the above definition of $y$ suggests that $y>0$. For the case $\alpha=0.4185$, this condition is fulfilled for $0<z<0.9165$.

Thus the solutions for $\tilde{M}$ and $\tilde{R}$ are obtained in terms of $z$ as 
\begin{equation}\label{eq:Mf}
\tilde{M} = \left(\frac{6\sqrt{\gamma}\alpha^3}{\pi}\right)^2 \left\{\frac{y(z)}{z(1-z^3)^{4/3}}\right\}^{3/2}
\end{equation}
and
\begin{equation}\label{eq:Xf}
\tilde{R} = \left(\frac{6\sqrt{\gamma}\alpha^3}{\pi}\right) \sqrt{\frac{y(z)}{z(1-z^3)^2}} 
\end{equation}

These expressions hold good for large $\xi$ or small $z$.  As $z\rightarrow0$, we find $\tilde{M}\sim z^{-3/2}$ and $\tilde{R}\sim z^{-1/2}$. Since $z\sim\xi^{-1}$, it follows that $\tilde{M}\sim \xi^{3/2}$ and $\tilde{R}\sim \xi^{1/2}$. Thus this simple treatment suggests that both mass and radius increase unboundedly with indefinite increase in the Fermi momentum. 

\subsection{Asymptotic solutions} 

In the limit $\xi\rightarrow0$, that is for low values of $\xi$, we see that $\nn(\xi)\sim3\xi^3/8$. We thus obtain 
\begin{equation}\label{eq:CHSN}
\frac{1}{x^2}\frac{d}{dx}\left(x^2\frac{d\sqrt{1+\xi^2}}{dx}\right)+\frac{3}{8}\xi^3 = 0.
\end{equation}
Letting $\sqrt{1+\xi^2}=c_0 \phi(x)$ and using the new dimensionless radius defined by $\eta=(\sqrt{3}c_0/2\sqrt{2})x$, the above equation becomes
\begin{equation}\label{eq:Chandra}
\frac{1}{\eta^2}\frac{d}{d\eta}\left(\eta^2\frac{d\phi}{d\eta}\right)+\left(\phi^2-\frac{1}{c_0^2}\right)^{3/2} = 0
\end{equation}
where $c_0$ is related to the central value $\xi_c$ by $c_0=\sqrt{1+\xi_c^2}$, when $\phi(0)=1$. Chandrasekhar \cite{Chandrasekhar1939} obtained the above equation for the case of ideal degenerate white dwarfs. For small values of $\xi_c$ we thus expect the mass-radius relation obtained with GUP equation of state to be the same as that obtained by Chandrasekhar. Since the above equation does not involve the parameter $\alpha$, this indicates that the effect of quantum gravity is imperceptible at low densities. 

On the other hand, for the asymptotic limit $\xi\rightarrow\infty$, the leading order term in $\nn(\xi)$ becomes a constant. Thus Eq.(\ref{eq:CHSR}) behaves like 
\begin{equation}\label{eq:CHSI}
\frac{1}{x^2}\frac{d}{dx}\left(x^2\frac{d\xi}{dx}\right)+\frac{\pi}{2\alpha^3}= 0.
\end{equation}

Putting $\xi/\xi_c=\phi(\zeta)$ in the above equation and redefining the dimensionless radius $\zeta=\pi x/2\alpha^3$, we get the zeroth order Lane-Emden equation whose solution is well known, namely, $\phi(\zeta)=c_1-\frac{c_2}{\zeta}-\frac{\zeta^2}{6}$, giving
\begin{equation}\label{eq:CHSsolu}
\frac{\xi(\zeta)}{\xi_c}=c_1-\frac{c_2}{\zeta}-\frac{\zeta^2}{6}.
\end{equation}
The undetermined constants can be fixed using the boundary condition $\xi(0)=\xi_c$. Accordingly we get $c_1=1$ and $c_2=0$, so that 
\begin{equation}\label{eq:CHSsoluR}
\xi(x)=\xi_c\left\{1-\frac{\pi}{12\xi_c\alpha^3}x^2\right\}. 
\end{equation}

Similarly, the asymptotic behavior of the mass can be obtained from the integral expression of Eq.~(\ref{eq:HS2}), namely, 
\begin{equation}\label{eq:totalM}
M=4\pi\int_0^{R} \rho(r) r^2 dr = 4\pi K \left(\frac{\mu_e m_u}{m_e c^2}\right) \int_0^{R} \nn(\xi) r^2 dr, 
\end{equation}
so that
\begin{equation}\label{eq:DMR}
\tilde{M}=\int_0^{\tilde{R}} \nn(\xi) x^2 dx.
\end{equation}

Since in the limit $\xi\rightarrow\infty$, $\nn(\xi)$ approaches the constant value $\pi/2\alpha^3$,  and since it is nearly true in the central region, we can define a core of approximately constant density. This is because of the fact that $\nn(\xi)$ already becomes $99.9\%$ of $\pi/2\alpha^3$ when $\alpha\xi \approx10.88$. Above this value, $\nn(\xi)\approx\pi/2\alpha^3$. Thus we can calculate the mass of the core $\tilde{M}_{\rm{core}}$  from the asymptotic limit $\nn(\xi)$ tending to $\pi/2\alpha^3$. This approximation will be no longer valid in the outer region. Assuming that the core extends up to a radius $\tilde{R}_{\rm{core}}$, we get $ \tilde{M}_{\rm core}= \nn(\infty) \tilde{R}_{\rm core}^3/3$. 

An estimate for the pressure inside the core can be obtained as
\begin{equation}\label{eq:PHuangcore}
P=\frac{\gamma}{4\pi}\frac{GM_{\rm core}^2}{R_{\rm core}^4}
\end{equation}
where $\gamma$ is a constant of order unity. Thus using Eq.~(\ref{eq:Pinfinity}) we obtain
\begin{equation}\label{eq:HScore}
K\left\{\nn(\infty)\xi_c+B(\alpha)\right\} = \frac{\gamma}{4\pi}\frac{GM_{\rm core}^2}{R_{\rm core}^4} 
\end{equation}

Converting mass and radius into dimensionless quantities, we obtain
\begin{equation}\label{eq:HScoreD}
\frac{\tilde{M}_{\rm core}}{\tilde{R}^2_{\rm core}} =  \frac{1}{\sqrt{\gamma}} \left\{\frac{\pi}{2\alpha^3}\xi_c+B(\alpha) \right\}^{1/2}
\end{equation}

From Eq.~(\ref{eq:CHSsoluR}), we can identify the radius of the core as $\xi({\tilde{R}_{\rm core}})=\xi_s$, so that 
\begin{equation}\label{eq:Xcoref}
\tilde{R}_{\rm core} = \left(\frac{12\alpha^3}{\pi}\xi_c\right)^{1/2} \sqrt{1-\frac{\xi_s}{\xi_c}}
\end{equation}
where $\xi_s$ is the value of $\xi$ such that $\nn(\infty)$ drops to $99.9\%$ of the central value $\nn(\infty)$. It is noticed from the $N(\xi)$ curve (Figure \ref{figure:1}) that the density reaches approximately constant values [$99.9\%$ of $\nn(\infty)$] for values of $\xi$ such that $\alpha\xi\geq10.88$, so that we can take $\xi_s=10.88/\alpha$. 

The above two relations lead to the mass of the core as 
\begin{equation}\label{eq:Mcoref}
\tilde{M}_{\rm core} =  \left(\frac{12\alpha^3}{\pi}\xi_c\right)    \frac{1}{\sqrt{\gamma}} \left\{\frac{\pi}{2\alpha^3}\xi_c+B(\alpha) \right\}^{1/2}       \left(1-\frac{\xi_s}{\xi_c}\right)
\end{equation}

Thus for $\xi_c\gg\xi_s$, $\tilde{M}_{\rm core}\sim\xi_c^{3/2}$ and $\tilde{R}_{\rm core}\sim\xi_c^{1/2}$. This implies that the mass and radius of white dwarfs increase unboundedly with unbounded increase in the central Fermi momentum. These results, including the scaling exponents with respect to the central Fermi momentum, are consistent with the results of Rashidi \cite{Rashidi2016}. They are also consistent with our previous heuristic analysis. In other words, the core dictates the mass-radius relation for excessively large central values of Fermi momentum $\xi=p_{\ff}/m_ec$. 

\subsection{Exact solution}
Now we shall consider the full equation of state (instead of asymptotics) in the Newtonian gravity.  Thus, after substituting Eqs. (\ref{eq:rho}) and (\ref{eq:P}) and using the definitions $m=M_0 u$ and $r=R_0 x$ in Eqs. (\ref{eq:HS1}) and (\ref{eq:HS2}), we obtain 
\begin{equation}\label{eq:HS1D}
\frac{d\xi}{dx}=-\frac{\nn(\xi)}{\pp'(\xi)}\frac{u(x)}{x^2}
\end{equation}
and 
\begin{equation}\label{eq:HS2D}
\frac{du}{dx}= \nn(\xi)x^2
\end{equation}
Since  $\pp'(\xi)=\frac{\xi}{\sqrt{1+\xi^2}}\nn(\xi)$,  Eq.~(\ref{eq:HS1D}) reduces to 
\begin{equation}\label{eq:HS1DS}
\frac{d\xi}{dx} =-\frac{\sqrt{1+\xi^2}}{\xi}\frac{u(x)}{x^2}  
\end{equation}
Using Eq.~(\ref{eq:curlyN}) in Eq.~(\ref{eq:HS2D}),  we obtain 
\begin{equation}\label{eq:HS2DS}
\frac{du}{dx}=\frac{1}{\alpha^3}\left\{\tan^{-1}(\alpha\xi)-\frac{\alpha\xi(1-\alpha^2\xi^2)}{(1+\alpha^2\xi^2)^2}\right\}x^2
\end{equation}

\begin{figure}
\centering
\includegraphics[width=8cm]{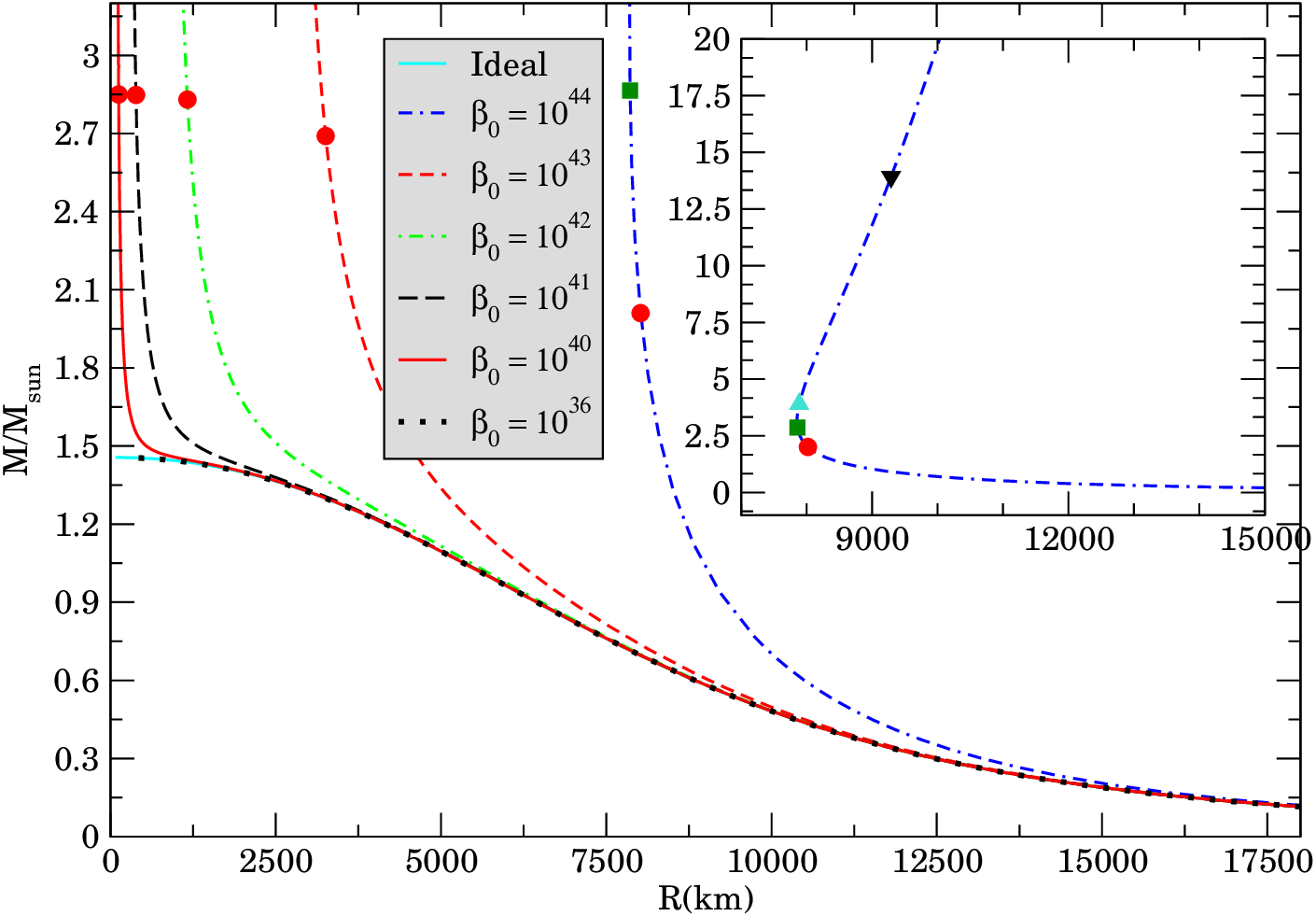}
\caption{\small Mass-radius curve for Helium white dwarfs with GUP equation of state for different values of $\beta_0$. The inset displays the mass-radius curve for $\beta_0=10^{44}$ up to very large values of the Fermi momentum$\xi_c$. The marked points on the mass-radius curve correspond to $\alpha\xi_c=1.00$ (solid circles), $1.25$ (solid squares), $1.50$ (solid triangle), $3.00$ (solid inverted triangle).}
\label{figure:4}
\end{figure}

We integrate the above two equations, namely, Eqs.~(\ref{eq:HS1DS}) and (\ref{eq:HS2DS}),  simultaneously with the boundary condition $\xi(0)=\xi_c$ and $u(0)=0$ until the surface defined by $\xi(\tilde{R})=0$ is reached. The result of the numerical integration for different values of $\xi_c$ and $\beta_0$ are shown in Figure \ref{figure:4}. For comparison, Chandrasekhar's  ideal degenerate case is also shown in the same figure. 

We see that, for large values of $\beta_0$,  the mass-radius relation departs more strongly from the ideal case than for smaller values. This is not a surprise since we expect the effect of GUP to be stronger for large values of $\beta_0$. Further, we note that the Chandrasekhar limit does not exist for any non-zero value of $\beta_0$ because the effect of quantum gravity takes over at large values of Fermi momentum $\xi$. (One can truly reproduce Chandrasekhar's limiting mass by completely neglecting quantum gravity effects, by setting $\beta_0=0$). Since the value of $\beta_0$ has a lower bound of $10^4$, $\alpha$ can be as low as $4.1854\times10^{-21}$, so that the effect of GUP can be almost imperceptible.  However to  assess the GUP effect, we took an exaggerated value of $\beta_0$, namely,  $10^{44}$. This situation is shown in the inset of Figure \ref{figure:4}.

Thus, for the case of $\beta_0=10^{44}$  (or equivalently $\alpha=0.4185$), it is observed from the right-hand part of Figure \ref{figure:4} that the mass-radius curve coincides with Chandrasekhar curve for low vales of $\xi_c$. As $\xi_c$ is increased, the mass increases slowly and the radius decreases, reaching a minimum value $\sim7853$ km,  as can be seen in the inset of Figure \ref{figure:4}. On further increasing $\xi_c$, the mass and radius both increase boundlessly which is consistent with the asymptotic analysis presented in Section 3.2, where we found that there exists a core of uniform density whose mass and radius are determined by the central value $\xi_c$.This is unlike the ideal case where the radius decreases to zero and the mass increases and approaches the Chandrasekar limit for increasing $\xi_c$. Thus it suggests that quantum gravity effect plays a significant role in determining the mass-radius relation.

\section{Mass and radius with restrictions}

The above study suggests an unrestricted increase in the mass and radius with unrestricted increase in the central value of the Fermi momentum $\xi_c=p_{\ff c}/m_e c$. However, the above analysis was performed without any restrictions that may otherwise alter the situation. In this section, we consider this and perform detailed analyses of the equations of hydrostatic equilibrium, namely Eqs.~(\ref{eq:HS1DS}) and (\ref{eq:HS2DS}),  by numerical means.  

\subsection{GUP Restriction} 

 As we discussed in Section 2.3, the GUP factor $f=(1+\beta p^2)^{-3}$ imposes a severe restriction on the availability of momentum states as it effectively puts an ultraviolet cutoff around $p_{\rm max}\sim\beta^{-1/2}$ (or equivalently $\xi_{max}\sim\alpha^{-1}$). Thus it may not be appropriate to consider the central values for the Fermi momentum much higher than $p_{\rm max}$. Consequently, as discussed earlier in Section 2.3, we shall take four different cases corresponding to $(\sqrt{\beta}p)_{\rm max}=(\alpha\xi)_{\rm max}=1.00, 1.25, 1.50$, and $3.00$. For these cases, $f$ reduces to $0.125, 0.059, 0.029$ and  $0.001$, respectively, signifying the fact that momentum states corresponding to the higher values of $\alpha\xi$ are scarcely available . We shall thus consider the central values of Fermi momentum such that $p_{\ff c} \leq p_{\rm max}$ (or equivalently $\xi_c \leq \xi_{\rm max}$) for different choices of $\beta_0$ (or $\alpha$). Moreover, as indicated in Section 1, there have been suggestions for lower and upper bounds for the parameter $\beta_0$, the extreme limits being $\beta_0>10^4$ and $\beta_0<10^{50}$. These bounds are equivalent to $\alpha>4.1854\times10^{-21}$ and $\alpha<4.1854\times10^{2}$. Since the upper bound $10^{50}$ is less reliable compared to the other upper bounds such as $10^{36}$ or $10^{34}$, we shall therefore consider cases when $\beta_0$ value is decreased through these latter values for which $\alpha$ becomes very small. It will be noticed that $\alpha$ become so small that we would even consider the limit $\alpha\rightarrow0$.
 
It may however be noted that the equation of state connecting $\rho(\xi)$ and $P(\xi)$ via Eqs.~(\ref{eq:rho}) and (\ref{eq:P}) reduces to the ideal equation of state for the limiting case $\alpha\rightarrow0$. Consequently, for the ideal case ($\alpha=0$), we would expect the Chandrasekhar limit. This is indeed the case  when we solve Eqs.~(\ref{eq:HS1DS}) and (\ref{eq:HS2DS})  for the case $\alpha=0$. This can be seen by taking the limit $\alpha\rightarrow0$ first in Eqs.~(\ref{eq:HS1DS}) and (\ref{eq:HS2DS}), that results in 
\begin{equation}\label{eq:HS1N}
\frac{d\xi}{dx}=-\frac{\sqrt{1+\xi^2}}{\xi}\frac{u(x)}{x^2}
\end{equation}
and 
\begin{equation}\label{eq:HS2N}
\frac{du}{dx}=\frac{8}{3}\xi^3 x^2
\end{equation}
which are exactly the same as those in the ideal case ($\alpha=0$). 

On the other hand, the $\alpha=0$ case is different from the case of a low value of $\alpha$, however small. We thus expect quite different solutions from the ideal case from Eqs.~(\ref{eq:HS1DS}) and (\ref{eq:HS2DS})  for low values of $\alpha$  and hence the limit $\alpha\rightarrow0$ does not coincide with the solution of Eqs.~(\ref{eq:HS1N}) and (\ref{eq:HS2N}). This can be seen by employing the transformations $\tilde{\xi}=\alpha \xi$ and $\tilde{x}=x/\alpha$ in Eqs.~(\ref{eq:HS1DS}) and (\ref{eq:HS2DS}), giving 

\begin{equation}\label{eq:HS1DT}
\frac{d\tilde{\xi}}{d\tilde{x}} =-\frac{\sqrt{\alpha^2+\tilde{\xi}^2}}{\tilde{\xi}}\frac{u(\tilde{x})}{\tilde{x}^2}  
\end{equation}
and 
\begin{equation}\label{eq:HS2DT}
\frac{du}{d\tilde{x}}=\left\{\tan^{-1}\tilde{\xi}-\frac{\tilde{\xi}(1-\tilde{\xi}^2)}{(1+\tilde{\xi}^2)^2}\right\}\tilde{x}^2.
\end{equation}
Thus taking the limit $\alpha\rightarrow0$ in Eqs.~(\ref{eq:HS1DT}) and (\ref{eq:HS2DT}) yield equations quite different from Eqs.~(\ref{eq:HS1N}) and (\ref{eq:HS2N}). Consequently, these differential equations do not reduce to the ideal differential equations even for very small values of $\alpha$, that is $\alpha\rightarrow0$.  

\begin{figure}
\centering
\includegraphics[width=9cm]{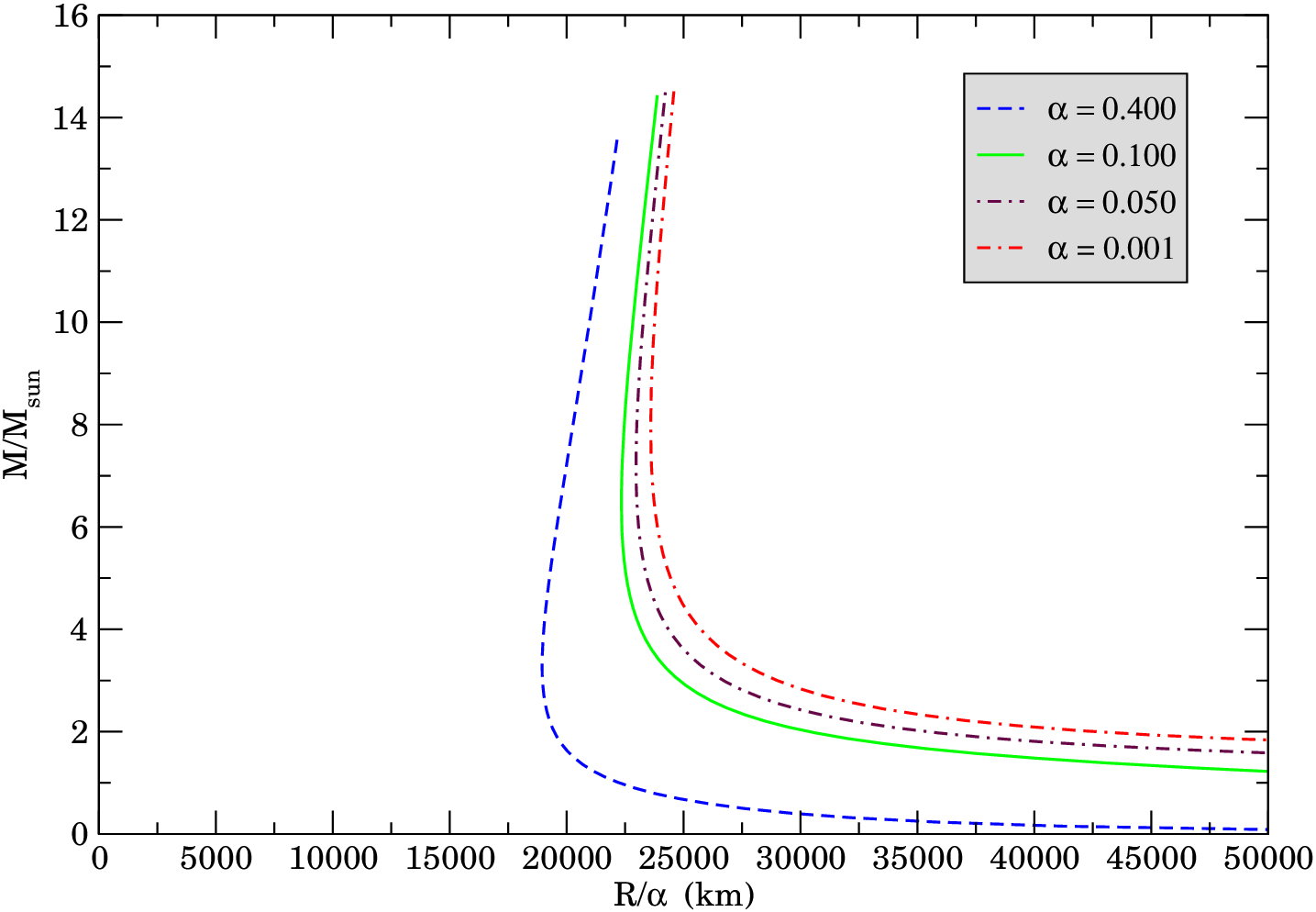}
\caption{\small Mass versus $R/\alpha$ for different values of $\alpha$. Each curve is obtained for central values of $\tilde{\xi}_c$ running from $3.00$ to $0.10$.}
\label{figure:5}
\end{figure}

We solve the above two differential equations, namely,  Eqs.~(\ref{eq:HS1DT}) and (\ref{eq:HS2DT}), starting with $\tilde{\xi}_c = 3.00$, for different decreasing values of $\alpha$. Mass versus the scaled radius $R/\alpha$  are shown in Figure \ref{figure:5} for representative values of $\alpha$, namely, $0.4, 0.1, 0.05$ and $0.01$. As we can see, the curves do not collapse, which is a consequence of the presence of $\alpha$ in Eq.~(\ref{eq:HS1DT}). The approach of the curves in the limit $\alpha\rightarrow0$ can also be seen in Figure \ref{figure:5}. In all cases, including the $\alpha=0$ case, we see that the mass of the white dwarf can be as high as $\sim14$ $M_{\odot}$.

Table \ref{table:3} shows the masses and radii for different central values, namely, $\tilde{\xi}_c=\alpha\xi_c=1.00, 1.25, 1.50$ and $3.00$. For each case, the parameter $\alpha$ is decreased down to $10^{-6}$ (which corresponds to $\beta_0=5.7085\times10^{32}$). It is apparent that, as $\alpha\rightarrow0$, the radius approaches zero, while the mass approaches a value that depends weakly on the choice of $\alpha$  but strongly on the the value of $\tilde{\xi}_c=\sqrt{\beta}p_{\ff c}$. This indicates that the maximum value of the allowed momentum plays a dominant role in determining the mass of the white dwarf. Moreover the limiting mass values so obtained do not seem to yield the Chandrasekhar limit even in the limit $\alpha\rightarrow0$.

\begin{table*}
\centering
\small
\caption{\small Four choices of $\tilde{\xi}_c$ and the corresponding value of mass and radius as $\beta_0$ decrease from $7.94292\times10^{43}$ to $5.70857\times10^{32}$.}
\label{table:3}
\vspace{0.1cm}
\begin{tabular}{cccccccccc}
\hline
\hline
\multirow{2}{*}{$\beta_0$} & \multirow{2}{*}{$\alpha$} & \multicolumn{2}{c}{$\tilde{\xi}_c=1.00$} &  \multicolumn{2}{c}{$\tilde{\xi}_c=1.25$} & \multicolumn{2}{c}{$\tilde{\xi}_c=1.50$} &\multicolumn{2}{c}{$\tilde{\xi}_c=3.00$} \\
\cmidrule(r){3-4} \cmidrule(r){5-6} \cmidrule(r){7-8} \cmidrule(r){9-10}
  &  & $R$ (km) & $M$ (M$_{\odot}$) & $R$ (km) & $M$ (M$_{\odot}$) & $R$ (km) & $M$ (M$_{\odot}$)& $R$ (km) & $M$ (M$_{\odot}$)\\
\hline   
$7.94292\times10^{43}$ &$0.373015$ & $7380.5923$ & $2.1168$ & $7193.0451$ & $2.9803$ & $7186.8669$ & $4.0253$   & $8364.3115$   & $14.1001$\\
$6.30929\times10^{42}$ &$0.105130$ & $2673.1655$ & $2.7424$ & $2472.1223$ & $3.5955$ & $2377.6702$ & $4.6455$   & $2517.7272$   & $14.8649$\\
$3.16223\times10^{41}$ &$0.023536$ & $675.0628 $ & $2.8436$ & $604.2194$  & $3.6820$ & $569.0310$  & $4.7251$   & $576.9071 $   & $14.9451$\\  
$9.99812\times10^{39}$ &$0.004185$ & $124.3514 $ & $2.8498$ & $110.1106$  & $3.6871$ & $103.0309$  & $4.7297$   & $103.1906 $   & $14.9494$\\
$3.97865\times10^{37}$ &$0.000264$ & $7.9054$    & $2.8500$ & $6.9834$    & $3.6872$ & $6.5253$    & $4.7298$   & $6.5187   $   & $14.9495$\\
$1.64978\times10^{35}$ &$0.000017$ & $0.4990$    & $2.8500$ & $0.4407$    & $3.6872$ & $0.4118$    & $4.7298$   & $0.4113 $     & $14.9495$\\
$5.70857\times10^{32}$ &$0.000001$ & $0.0396$    & $2.8500$ & $0.0350$    & $3.6872$ & $0.0327$    & $4.7298$   & $0.0327 $     & $14.9495$\\
\hline
\hline
\end{tabular}
\end{table*}

\begin{figure}
\centering
\includegraphics[width=9cm]{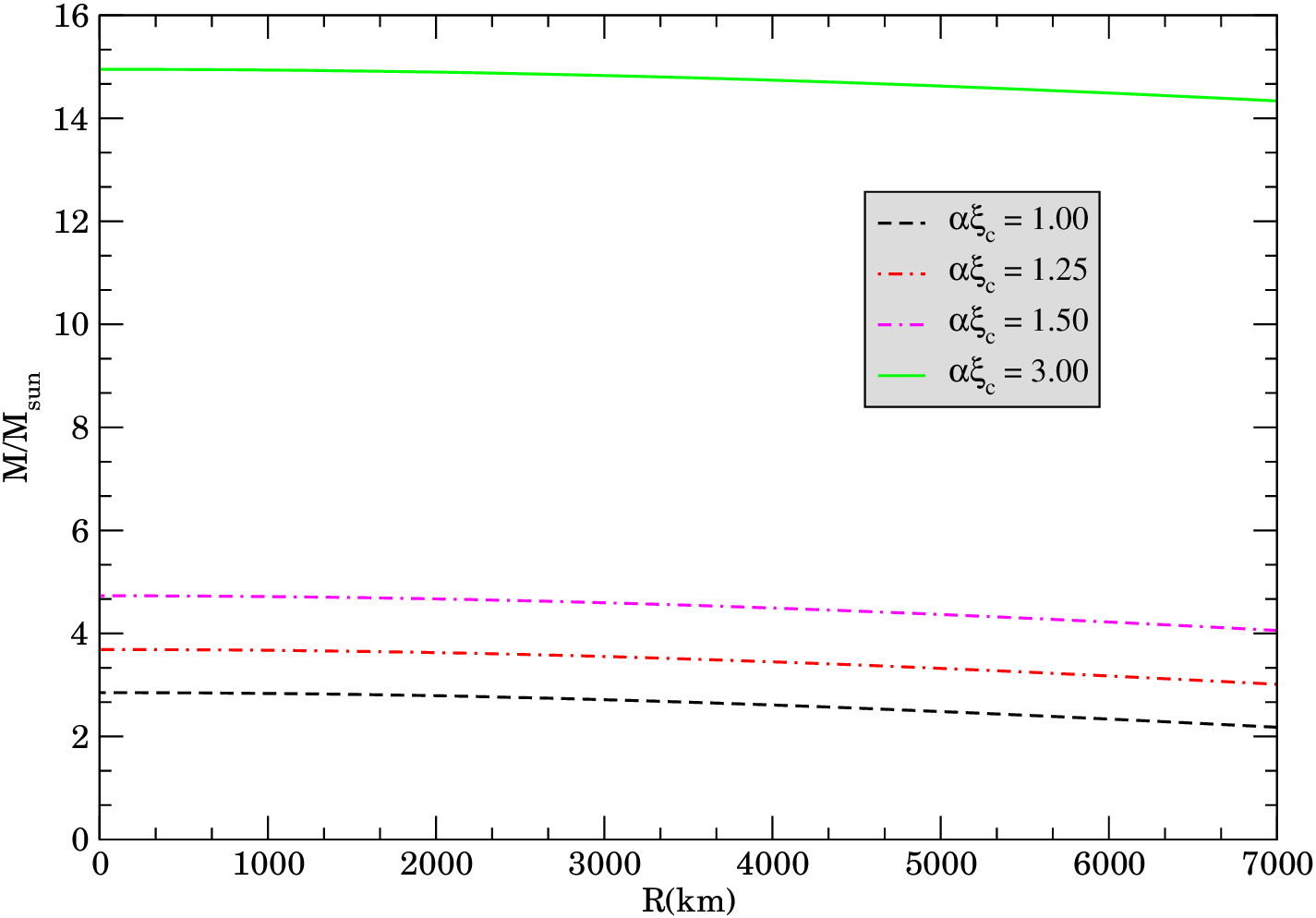}
\caption{\small Mass-radius curves for four choices of $\alpha \xi_c$ (or $\tilde{\xi}_c$).  The value of $\alpha$ decrease from right to left. A few representative numerical values are displayed in Table \ref{table:3}.}
\label{figure:6}
\end{figure}

In Figure~\ref{figure:6}, we display the mass-radius relation holding $\tilde{\xi}_c = \alpha\xi_c$ at four constant values, namely, $\tilde{\xi}_c=1.00, 1.25, 1.50$ and $3.00$. Each curve corresponds to solutions of Eqs.~(\ref{eq:HS1DT}) and (\ref{eq:HS2DT}) with the parameter $\alpha$ decreasing continuously from right to left (down to $10^{-6}$). We see that, as $\alpha\rightarrow0$, the mass approaches different constant values depending on the value of $\tilde{\xi}_c$, whereas the radius approaches zero. These limiting mass values are higher than the Chandrasekhar limit.

\subsection{Restriction due to Neutronization} 

Inverse $\beta$-decay ($^A_ZX + e\longrightarrow ^{\ \ A}_{Z-1}\!\!Y +\nu_e$), or neutronization, sets in at a sufficiently high density when the condition on the Fermi energy $E_F\geqslant \varepsilon_Z$ is satisfied, where  $\varepsilon_Z$ is the difference in binding energies of the parent and daughter nuclei. The threshold density $\rho_{N}$ for the ideal degenerate case was calculated by Salpeter \cite{Salpeter1961} by setting $E_F=\varepsilon_Z$.  Following Salpeter, we obtain the neutronization threshold Fermi momentum $p_N$ as 
\begin{equation}\label{eq:}
\xi_N=\frac{p_N}{m_e c}=\sqrt{\frac{\varepsilon_Z^2 + 2 m_e c^2 \varepsilon_Z}{m_e^2 c^4}}.
\end{equation}
The inverse $\beta$-decay energy  $\varepsilon_Z$ for Helium can be found from Table \rom{2} of  Rotondo et al. \cite{Rueda2011}. The $\beta$-decay energies $\varepsilon_Z$ for various isotopes of different elements were obtained from least square fits to the experimental data by Wapstra and Bos \cite{Wapstra1977}. For $^4_2$He$, \varepsilon_Z=20.596$ MeV and the corresponding $\xi_N=41.2932$. 

Since in the GUP framework, the equation for hydrostatic equilibrium are expressed by  Eqs.~(\ref{eq:HS1DT}) and (\ref{eq:HS2DT}) in term of $\tilde{\xi}=\alpha \xi$, the neutronization threshold value $\tilde{\xi}_N=\alpha\xi_N$ takes different values for different choices of $\alpha$, or equivalently $\beta_0$. Consequently we solve Eqs.~(\ref{eq:HS1DT}) and (\ref{eq:HS2DT}) numerically for different values of $\alpha$ taking the central value as $\tilde{\xi}_N$. Table \ref{table:4} gives the values of $\tilde{\xi}_N$ corresponding to different deceasing values of $\alpha$. The corresponding mass and radius of the white dwarfs are shown in the last two columns. One may notice that both mass and radius approach finite limiting values. For excessively large values of $\beta_0$, such as $10^{44}$ (and higher), both mass and radius increase to very high values. However, as the $\beta_0$ value is decreased to $10^{40}$, the mass and radius are approximately $1.48$ M$_{\odot}$ and $600$ km.

 Although a high value of the upper bound such as $10^{50}$ has been suggested on the basis of inaccuracy in STM measurements of Landau levels, this upper bound is not reliable because STM measurements are accurate only up to $1$ part in $10^3$. On the other hand, the accuracy of Lamb shift in hydrogen atom is at the level of $1$ part in $10^{12}$, and therefore the corresponding upper bound of $10^{36}$ is more reliable. We see from Table \ref{table:4} that the mass and radius values are approximately $1.45$ M$_{\odot}$ and $600$ km for the legitimate choice of $\beta_0=10^{36}$.
 
 We also see from Table \ref{table:4} that as $\beta_0$ value is decreased to $10^{34}$, $\alpha$ reduces to a very small number $\sim10^{-6}$ and $\tilde{\xi}_N$ reduces to $\sim10^{-4}$, so that $\alpha\ll\tilde{\xi}_N$. On further decreasing $\beta_0$ to $10^{32}$, we get $\alpha\sim10^{-7}$ and $\tilde{\xi}_N\sim10^{-5}$, and it becomes very difficult to solve Eqs.~(\ref{eq:HS1DT}) and (\ref{eq:HS2DT}) numerically. Thus for $\beta_0\leq10^{32}$, we solve Eqs.~(\ref{eq:HS1DT}) and (\ref{eq:HS2DT}) approximately by assuming $\alpha\ll\tilde{\xi}_N$, obtaining  
\begin{equation}\label{eq:HS1DN}
\frac{d\tilde{\xi}}{d\tilde{x}} =-\frac{u(\tilde{x})}{\tilde{x}^2}  
\end{equation}
and 
\begin{equation}\label{eq:HS2DN}
\frac{du}{d\tilde{x}}=\frac{8}{3}\tilde{\xi}^3\tilde{x}^2,
\end{equation}
since $\tilde{\xi}_N$ becomes very small as indicated above. Combining Eqs.~(\ref{eq:HS1DN}) and (\ref{eq:HS2DN}), and defining a new dimensionless radius $\eta=\sqrt{\frac{8}{3}}\tilde{\xi}_c\tilde{x}$, and defining $\theta=\tilde{\xi}/\tilde{\xi}_c$, we obtain 
\begin{equation}\label{eq:Emden3}
\frac{1}{\eta^2} \frac{d}{d\eta}  \left(\eta^2 \frac{d\theta }{d\eta}      \right)+\theta^3 = 0
\end{equation}
with boundary conditions $\theta(0)=1$ and $\theta(\eta_R)=0$, where $\eta_R$ corresponds to the radius of the white dwarf. Eq.~(\ref{eq:Emden3}) is the Lane-Emden equation of index 3, whose numerical solutions are already known. From Eq.~(\ref{eq:HS2DN}) we obtain the mass of the white dwarf in terms of the new coordinate as 
\begin{equation}\label{eq:}
\tilde{M}_N = -\sqrt{\frac{3}{8}} \eta_R^2 \left(\frac{d\theta}{d\eta}\right)_{ \eta=\eta_R}
\end{equation}

Numerical solution for  the Lane-Emden equation of index 3 can be found in Weinberg \cite{Weinberg1972}. For $n=3$, $-\eta_R^2 \left(\frac{d\theta}{d\eta}\right)_{ \eta=\eta_R}=2.01824$, thus giving
\begin{equation}\label{eq:}
M_N=M_0\tilde{M}_N=1.4563 \ \rm{M}_{\odot}.
\end{equation}

We also obtain the radius of the white dwarf as 
\begin{equation}\label{eq:neutroR}
\tilde{R}_N=\sqrt{\frac{3}{8}}\frac{\alpha}{\tilde{\xi}_N}\eta_R = \sqrt{\frac{3}{8}}\frac{1}{\xi_N}\eta_R
\end{equation}
where, for $n=3$, $\eta_R=6.89685$, thus  
\begin{equation}\label{eq:}
 R_N=R_0\tilde{R}_N=648.81 \ \rm {km}.
\end{equation}

We note that the factors of $\alpha$ in the expression for radius given by Eq.(\ref{eq:neutroR}) get cancel`ed making it independent of $\alpha$. This appears to be the feature when the $\alpha$ value is low. These mass and radius values are shown in the last row of Table \ref{table:4} that appears to be consistent with the numerical solutions given in the previous few rows. 

Due to neutronization in the core, the star is expected to collapse to form a neutron star. Since we do not include the effect of neutronization in the equation of state, this situation falls outside the scope of our analysis.  

\begin{table}
\centering
\caption{\small Mass and radius for different values of $\beta_0$ when the central Fermi momentum is taken to be the neutronization threshold $\xi_N=41.2932$.The third entry, $\beta_0=7.6\times10^{41}$, corresponding to $\tilde{\xi}_N\approx1.5$, has been included to make a comparison with the GUP restriction $\tilde{\xi}_{\rm max}=1.5$. The last entry, $\beta_0<10^{32}$, shows approximate estimates in the limit $\alpha\rightarrow0$.}
\label{table:4}
 \vspace{0.1cm}
\begin{adjustbox}{width=0.45\textwidth}
\begin{tabular}{ccccc}
\hline
\hline
$\beta_0$   &  $\alpha$  & $\tilde{\xi}_N$    &$R_{\rm N}$ (km) & $M_{\rm N}$ (M$_{\odot}$) \\
\hline
$10^{44}$                        & $0.4185393$                      & $17.28281$                    & $21375.47$         & $294.7258$\\  
$10^{42}$                        & $0.0418539$                      & $1.72828$                      & $972.65$             & $5.8439$\\  
$7.6\times10^{41}$         & $0.0364874$                       & $1.50668$                     & $871.17$             & $4.7494$\\  
$10^{41}$                       & $0.0132354$                       & $0.54653$                      & $614.74$             & $1.8457$\\ 
$10^{40}$                       & $0.0041854$                       & $0.17283$                      & $601.33$             & $1.4898$\\
$10^{38}$                       & $0.0004185$                       & $0.01728$                      & $601.22$             & $1.4521$\\
$10^{36}$                       & $0.0000419$                       & $0.00173$                      & $601.24$             & $1.4518$\\  
 $10^{34}$                      & $4.2\times10^{-6}$              & $0.00017$                      & $612.47$             & $1.4518$\\
 $\beta_0<10^{32}$        & $\alpha\ll1$                         &                                         & $648.81$               & $1.4563$\\
 \hline
 \hline
\end{tabular}
\end{adjustbox}
\end{table}

\section{Discussion and Conclusion}

In this paper, we considered the effect of generalized uncertainty principle on the mass-radius relationship of Helium white dwarfs. The generalized uncertainty is believed to have its origin in quantum gravity and such uncertainty relations have been framed in string theory. The generalized uncertainty relation we have considered modifies the measure of the phase space integral by introducing a factor of $(1+\beta \textbf{p}^2)^{-3}$. As a consequence, the expressions for the density and pressure of the degenerate electron gas assume forms different from the ideal case and they now depend on the parameter $\beta$. We saw that the density now approaches a constant value whereas the pressure increases linearly in the region of ultra-high Fermi momentum. This is quite unlike the ideal case ($\beta=0$) where both these quantities increase boundlessly with respect to the Fermi momentum. In consequence, the equation of state connecting the density and pressure has a completely different behavior from the ideal case. Since the stability of a star is determined by the equation of state of the matter it contains, we expect a different mass-radius relation for white dwarfs from the ideal case. 

 Assuming the density to be uniform, we first carried out a heuristic analysis and derived an approximate mass-radius relation in a parametric form. This yielded the behaviour $M\sim\xi^{3/2}$ and $R\sim \xi^{1/2}$ for ultra-high values of the Fermi momentum. Thus this simple analysis suggested that the mass and radius increase boundlessly in the limit of high Fermi momentum. 

To obtain a better picture, we carried out an asymptotic analysis of the hydrostatic  equilibrium (incorporating density variation) in the limit of high Fermi momentum. This analysis suggested that a white dwarf with ultra-high central Fermi momentum $\xi_c$ will have a core of constant density determined by the Fermi momentum $\xi_s$ above which the density remains constant. Both the mass and radius of the core are determined by the onset value $\xi_s$ and the central value $\xi_c$. However, for arbitrarily high values of the central Fermi momentum $\xi_c\gg\xi_s$,  the core mass and radius acquire negligible contribution from $\xi_s$, yielding $M_{\rm core}\sim\xi_c^{3/2}$ and $R_{\rm core}\sim\xi_c^{1/2}$. The same behavior was obtained by Rashidi \cite{Rashidi2016} in the limit $\xi_c\rightarrow\infty$. These analyses  suggest that the mass and radius increase boundlessly with unbounded increase in the central Fermi momentum. 

To get a full picture of the mass-radius relationship, we obtained an exact solution of the hydrostatic equation of equilibrium by a numerical scheme. It was observed that, for $\beta_0=10^{36}$ (corresponding to $\alpha = 4.1854\times10^{-5}$), the mass-radius relation follows closely the original relation (without GUP) down to $\sim 100$ km. The same situation occurs for any value of $\beta_0<10^{36}$. For $\beta_0=10^{40}$  (corresponding to $\alpha =4.1854\times10^{-3}$), the mass-radius curve diverges for low values of radii. Thus for the case  $\beta_0=10^{36}$ also we expect the same kind of divergence to occur towards a much lower value of radius. The mass-radius relation can be analyzed in a better way by looking at the behavior for  large values for $\beta_0$. Thus, for the case $\beta_0=10^{44}$ (correspondingly to $\alpha=0.4185$), it is observed from the right hand part of Figure \ref{figure:4} that the mass-radius curve coincides with the Chandrasekhar curve for low values of $\xi_c$. As $\xi_c$ is increased, the mass increases slowly and the radius decreases, reaching a minimum value $\sim7853$ km as can be seen in the inset of Figure \ref{figure:4} (moving from right to left). On further increasing $\xi_c$, both the mass and radius increase boundlessly, which is consistent with our previous asymptotic analysis where we found that there exists a core of uniform density whose mass and radius are determined by the central value of $\xi_c$. Thus, this ultra-high region of Fermi momenta corresponds to a the behavior $M\sim\xi_c^{3/2}$ and $R\sim\xi_c^{1/2}$, which was also obtained by \cite{Rashidi2016}.

We also noted that the GUP factor $(1+\beta \mathbf{p}^2)^{-3}$ imposes a severe restriction on the allowed values of momentum as this factor effectively puts an ultraviolet cutoff $p_{\rm max}\sim\beta^{-1/2}$ in the momentum integral. Thus, it appears that the center of a white dwarf cannot have an arbitrarily high value of Fermi momentum.  To address this problem, we took four representative values, $(\sqrt{\beta}p_{\ff})_{\rm max} =1.00, 1.25, 1.50$, and $3.00$. Noting that the GUP parameter $\beta_0$ (or equivalently $\alpha$) can be as low as $10^{4}$ (or equivalently $\alpha$ as low as $4.1854\times10^{-21}$), we obtained the limiting mass in each case by decreasing the value of $\beta_0$ down to $10^{32}$ (or equivalently $\alpha\sim 10^{-6}$). We note that, for a particular choice of $(\sqrt{\beta}p_{\ff})_{\rm max}$ the mass appears to converge to a limiting value whereas the radius approaches zero, which can be seen already at $\beta_0\sim10^{32}$ when the $\beta_0$ value is gradually decreased from $\beta_0\sim10^{43}$. As demonstrated in Table \ref{table:3}, such GUP restrictions lead to different values of limiting mass that are higher than the Chandrasekhar limit. On the other hand, if the GUP restriction is relaxed, and the central value of $\sqrt{\beta}p_{\ff}$ is allowed to take arbitrarily high values, the mass increases boundlessly as demonstrated in Figure \ref{figure:5}. 

Since Inverse $\beta$-decay puts a \textit{realistic} restriction on the extremal values of the mass and radius of a white dwarf, we next considered the maximum value of the central Fermi momentum to be determined by the neutronization threshold. However, since the value of the parameter $\beta_0$ is presently unknown, we considered different values of $\beta_0$ ranging from $10^{44}$ to $10^{34}$. We solved the equations of hydrostatic equilibrium numerically with these $\beta_0$ values. For the case $\beta_0=10^{44}$, the mass and radius turn out to have large values, $\sim295$ M$_{\odot}$ and $\sim21375$ km. On the other hand, for $\beta_0=10^{42}$, the mass is $\sim6$ M$_{\odot}$ and radius is $\sim 970$ km, whereas for $\beta_0=10^{40}$ , we obtained $\sim1.5$ M$_{\odot}$ and $\sim 601$ km, respectively. Although an upper bound of $10^{50}$ for $\beta_0$ was suggested on the basis of STM measurements of Landau levels, this upper bound is not reliable because STM measurements have a low accuracy of $1$ part in $10^3$. On the other hand, measurements on Lamb shift in a Hydrogen atom with a much higher accuracy of $1$ part in $10^{12}$ yielded an upper bound of $10^{36}$. Consequently we take this more reliable upper bound for granted. We thus considered the value $\beta_0=10^{36}$ and performed the numerical calculations with the neutronization threshold $\xi_N$ as the central Fermi momentum. We found that the mass and radius now turn out to be $1.45$ M$_{\odot}$ and $601$ km, respectively. We also considered a lower value $10^{34}$ and the numerical solutions yielded almost the same values of mass and radius. On the other hand, when $\beta_0$ is decreased sufficiently from $10^{34}$ so that we could take $\alpha\ll 1$, the mass and radius are found to be approximately $1.46$ M$_{\odot}$ and $650$ km. Since in stable Helium white dwarfs, the central Fermi energy is expected to be lower than the neutronization threshold value of $20.596$ MeV, they are expected to have masses lower than $1.46$ M$_{\odot}$ and radii larger than $650$ km. This can be confirmed by solving Eqs.~(\ref{eq:HS1DS}) and (\ref{eq:HS2DS}) with central Fermi energy less than $20.596$ MeV and $\beta_0\leq10^{36}$. Moreover, for stable white dwarfs with core composition other than Helium the neutronization threshold value is much lover than $20.596$ MeV.  For example, for $^{12}_6\rm C$, $^{16}_{8}\rm O$, and $^{20}_{10}\rm Ne$, the neutronization threshold values are $13.370$ MeV, $10.419$ MeV, and $7.026$ MeV, \cite{Rueda2011, Wapstra1977}  respectively. Consequently white dwarfs with these compositions are expected to have central Fermi energies lower than these values. Since most white dwarfs occurring in Nature have their cores composed of these elements, they are expected to have masses much lower than $1.46$ M$_{\odot}$ and radii much larger than $650$ km. These conclusions are consistent with the observations of non-magnetic white dwarfs that are usually found to be in the mass range from $0.17$ M$_{\odot}$ \cite{Kilic2007} to $1.33$ M$_{\odot}$ \cite{Vennes1997, Vennes1999, Marsh1997, Kepler2007} with radii ranging from $0.0153$ $R_{\odot}$ ($10644$ km) to $0.0071$ $R_{\odot}$ ($4939$ km) \cite{Shipman1972, Shipman1977, Shipman1979, Bond2017}.

We conclude by noting that the above discussions based on inverse $\beta$-decay threshold suggest that the range of mass and radius values of white dwarfs observed in Nature are dictated by the neutronization threshold in the presence of the effects of quantum gravity. 

\section*{Acknowledgement}
Arun Mathew is indebted to the Ministry of Human Resource Development, Government of India, for financial support through a doctoral fellowship.  

\end{document}